\documentclass[conference]{IEEEtran}
\IEEEoverridecommandlockouts

\usepackage{cite}
\usepackage{amsmath,amssymb,amsfonts}
\usepackage{algorithmic}
\usepackage{graphicx}
\usepackage{url}
\usepackage{textcomp}
\usepackage{xcolor}
\usepackage{booktabs} 
\usepackage{multirow} 
\usepackage{array}    

\def\BibTeX{{\rm B\kern-.05em{\sc i\kern-.025em b}\kern-.08em
    T\kern-.1667em\lower.7ex\hbox{E}\kern-.125emX}}

\begin{document}

\title{Investigating Cultural Dimensions and Technological Acceptance:\\
The Adoption of Electronic Performance and Tracking Systems\\
in Qatar's Football Sector\\
{\footnotesize \textsuperscript{}}
\thanks{This research received no external funding.}
}

\author{\IEEEauthorblockN{Abdulaziz Al Mannai}
\IEEEauthorblockA{\textit{Carnegie Mellon University} \\
Doha, Qatar \\
abdulazm@andrew.cmu.edu}
}

\maketitle

\begin{abstract}
Qatar's football sector has undergone a substantial technological transformation through the implementation of Electronic Performance and Tracking Systems (EPTS). This paper examines the impact of cultural and technological factors on EPTS adoption, guided by Hofstede’s Cultural Dimensions Theory and the Technology Acceptance Model (TAM). An initial exploratory study was conducted with ten participants, and a subsequent expanded dataset included thirty stakeholders (players, coaches, and staff) from Qatari football organizations. We employed multiple regression analysis to investigate relationships between perceived usefulness, perceived ease of use, power distance, innovation receptiveness, integration complexity, and overall adoption. Results show that perceived usefulness, innovation receptiveness, and lower power distance significantly drive EPTS adoption, while ease of use is marginally significant and integration complexity is non-significant in this sample. These findings offer practical insights for sports technology stakeholders in Qatar and highlight the need to balance cultural considerations with technological readiness for effective EPTS integration.
\end{abstract}


\section{Introduction}

Electronic Performance and Tracking Systems (EPTS) have had an incredible effect on
professional football in Qatar, marking a global shift towards increasing athletic
performance with data-driven approaches \cite{isspf2023}. EPTS, with its wide array of
technologies such as sleep management software and player monitoring systems, has become
an indispensable way of filling informational voids about player performance in the field
\cite{knowledgenile2023}. Utilizing sophisticated camera-based and wearable technologies,
these systems enable comprehensive observations of player positions, physiological
parameters, team dynamics and player development \cite{fifa2021, tierney2021}. As such,
these systems enable football coaching and player development programs to approach football
games more strategically and intelligently \cite{fifa2021, tierney2021}.

Integration of EPTS technologies—long staples among elite international football clubs—
presents both unique challenges and prospects in Qatari football \cite{aughey2010}. The 
adoption and incorporation of EPTS are significantly influenced by Qatar's organizational
and cultural environment, with its distinctive values and practices \cite{hofstede1980}.
Furthermore, factors such as user-friendliness and system compatibility can strongly influence
EPTS adoption success in the region, aligning with the Technology Acceptance Model (TAM)
\cite{davis1989}.

Given the significant effect of cultural and technological factors on EPTS adoption in Qatar,
this research intends to investigate them thoroughly. It will analyze how cultural values,
such as collectivism and power distance as described by Hofstede \cite{hofstede1980}, as well 
as technological considerations like perceived usefulness and ease of use via the TAM
framework \cite{davis1989}, influence EPTS use and effectiveness within Qatar's football sector.

This investigation's research question is:
\emph{"How do specific cultural dimensions, as defined by Hofstede's Cultural Dimensions Theory, 
and technological factors (delineated in the Technology Acceptance Model), influence adoption
and effectiveness of electronic performance tracking systems within Qatar's football industry?"}

By conducting this investigation, this study provides an in-depth examination of Qatar's
EPTS adoption process, contributing scholarly insights into tech integration across various
organizational and cultural environments.

\section{Literature Review}

This literature review presents an in-depth analysis of Electronic Performance and Tracking 
Systems (EPTS) within football and the wider sports business, focusing on how technology, 
culture, and user perceptions shape adoption.

\subsection{EPTS Technological Advancements}
EPTS's cutting-edge technologies have revolutionized sports statistics. Lozano and Muyor (2021) 
discuss FIFA's efforts in setting global standards for EPTS, which validate data such as player 
position and speed. Schmid and Lames \cite{schmid2023} detail how EPTS errors across sports 
can be corrected through precise measurements of positions, speeds, and accelerations. Linke 
et al. \cite{linke2018} investigate the measurement accuracy of tracking technologies such as 
GPS and video systems, demonstrating how technological advancement has enhanced performance
monitoring more reliably.

\subsection{Cultural and Social Influences}
Cultural and social factors significantly impact EPTS adoption within sports environments. 
Ratten \cite{ratten2019} emphasizes the interplay of culture, society, and commercialization in 
sports technology innovation. Lee et al. \cite{lee2013} highlight how cultural differences 
influence technology acceptance, while Oc and Toker \cite{oc2022} use “context-awareness” to 
explain the complex adoption patterns in sports settings.

\subsection{User Perceptions and Practical Implementation}
To successfully integrate EPTS, it is critical to understand perceptions among professional
football stakeholders. Tierney \cite{tierney2021} highlights how prevalent wearable technology 
is among association football teams, with many leveraging EPTS to improve performance. Bitilis 
\cite{bitilis2021} explores perspectives of professional players and coaching staff regarding
EPTS utilization, offering valuable insights into real-world challenges. Kim et al. \cite{kim2023} 
discuss the combination of GNSS/IMU wearable EPTS as a major technological advance.

\subsection{Injury Prevention and Future Innovations}
EPTS also contributes to injury prevention and foreseeing technological advancements. Dunn 
et al. \cite{dunn2018} examine football injury scenarios where EPTS may help mitigate risks. 
Manu-Ks \cite{manuks2020} reviews AI’s current and potential uses in football, cautioning about
ethical considerations. Together, these studies paint a complex picture in which technology, 
culture, and practical usage intersect to shape the future of sports analytics.

\section{Theory}

\subsection{Hoftede Framework}
To answer my research question, I decided to conduct my study using Hofstede Framework
cultural adoption theory as the basis. Hofstede's Cultural Dimensions Theory provides a
framework for understanding cross-cultural variations in business practices and national cultural
differences. Put another way, this framework acts as a means for distinguishing among various
national cultures, highlighting cultural traits, analyzing how these may alter behavior patterns,
and facilitating communication within multiple contexts, including business and diplomacy.
Hofstede (1980) developed his cultural dimensions theory as part of a comprehensive survey he
conducted throughout the 1960s and 70s in relation to value disparities within various IBM
divisions \cite{hofstede1980}.

Hofstede's Cultural Dimensions Theory can bring significant innovation and relevance to the
football industry, particularly Qatari teams which feature multicultural composition and unique
national culture. Examining how power distance and collectivism influence perception and
utilization of EPTS for sports purposes is central to this research project.

Hofstede's framework offers us insight into how cultural differences impact adoption and usage
of EPTS, providing us with insight into how cultural dimensions such as individualism vs
collectivism or power distance may significantly impact how players and staff use and appreciate
EPTS systems. Though initially created for business settings, Hofstede's framework can also be
applied effectively when studying sports teams to assess cultural dynamics that may influence
technology adoption.

\begin{figure}[ht]
    \centering
    \includegraphics[width=0.55\linewidth]{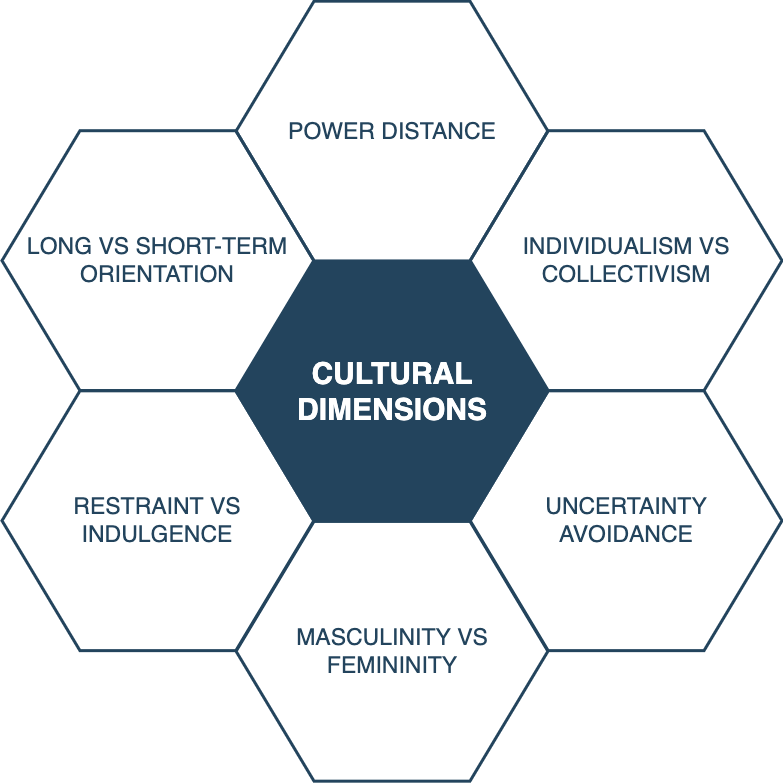}
    \caption{Hofstede Framework (Hofstede, 1980)}
    \label{fig:hofstede}
\end{figure}

\vspace{2mm}

\subsection{Technology Acceptance Model}
\begin{figure}[ht]
    \centering
    \includegraphics[width=0.5\linewidth]{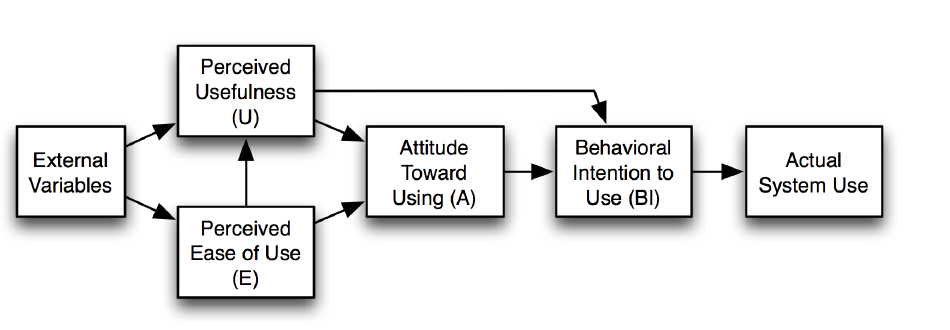}
    \caption{Technology Acceptance Model (Davis, 1989)}
    \label{fig:tam}
\end{figure}

Davis (1989) created the Technology Acceptance Model (TAM), an influential framework in
technology adoption. TAM highlights two primary influences that impact an individual's decision
to adopt new technology: perceived ease of use and perceived usefulness. Davis used TAM to
map patterns of computer use, as evidenced in Figure~\ref{fig:tam}. Davis' TAM seeks to identify
the fundamental determinants that contribute to computer acceptance, providing insight into user
behavior across a spectrum of end-user computing technologies and demographics. Perceived
usefulness refers to an individual's belief in the likelihood that employing a certain system, like
an online payment solution, will improve their performance. Conversely, perceived ease of use
refers to how much users anticipate that a system will be user-friendly \cite{davis1989}. External
variables in Lai's TAM framework can have an influence over these perceptions.

TAM is an industry standard model for understanding how individuals adapt new technologies,
particularly EPTS systems in football. When applied to EPTS specifically, TAM
provides insight into factors affecting acceptance and use by players, coaches, and other
stakeholders. Perceived utility and user friendliness are two hallmarks of TAM that play an
essential role in sports contexts. The model allows you to investigate whether perceived benefits
(such as performance improvement and injury prevention) and user friendliness of EPTS affect
its acceptance within the football industry. TAM's ability to incorporate external variables makes
it ideal for studying EPTS adoption. You can add sports industry-specific factors like team
dynamics, technical infrastructure, and organizational support to gain a better
understanding of their effect on technology acceptance.

Perceived usefulness and ease of use are critical in Qatar's football industry, determining how
players, coaches, and administrative bodies perceive and adopt EPTS
systems. Our model will explore this aspect within Qatari teams to better understand their
specific dynamics as well as inclination towards adopting such advanced systems.

\vspace{2mm}

\subsection{Research Model}

\begin{figure}[ht]
    \centering
    \includegraphics[width=0.8\linewidth]{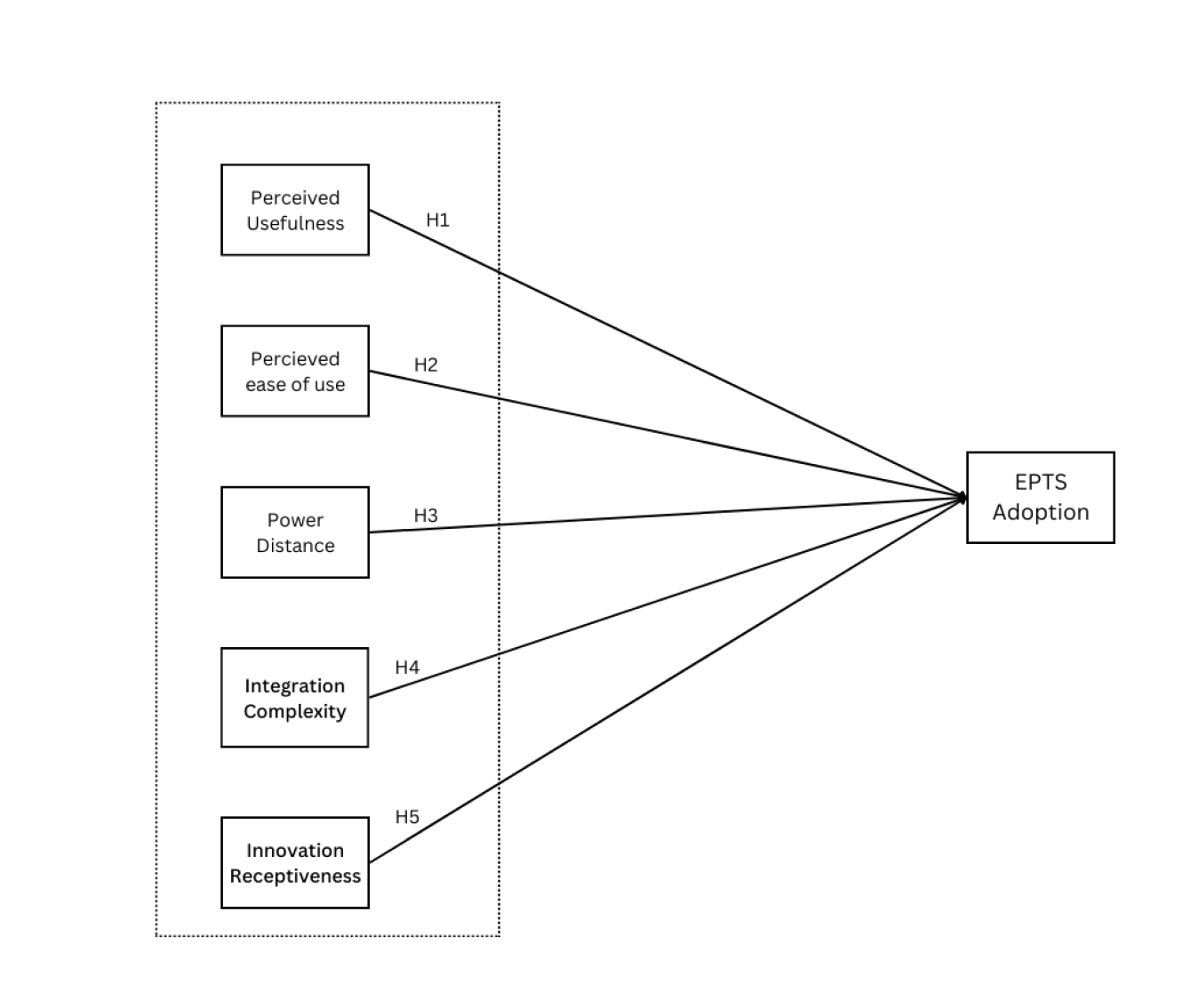}
    \caption{Proposed Research Model combining Hofstede and TAM constructs.}
    \label{fig:research_model}
\end{figure}

This study presents an in-depth research model (Fig.~\ref{fig:research_model}) to better
comprehend the adoption of EPTS within Qatar's football industry, drawing upon Hofstede's
Cultural Dimensions Framework and the Technology Acceptance Model (TAM). The model posits
that cultural factors can greatly impact Perceived Usefulness (PU) and Perceived Ease of Use
(PEU), shaping attitudes about EPTS acceptance and determining its eventual implementation
among players and staff.

\vspace{2mm}

\subsection{Hypothesis Development}

\paragraph{Perceived Usefulness (PU)}
PU is integral in EPTS adoption within football industries. Past research \cite{davis1989} 
suggests that coaches/players who perceive technology as performance-enhancing show greater
adoption intention. \textbf{H1:} PU is positively associated with EPTS 
.

\paragraph{Perceived Ease of Use (PEU)}
Integration complexity often depends on user-friendliness, especially in Qatar’s sports 
context \cite{venkatesh2008}. \textbf{H2:} PEU is positively associated with EPTS adoption.

\paragraph{Power Distance (PD)}
Qatar’s hierarchical structures (Hofstede \cite{hofstede1980}) can hamper open communication
about new tech. \textbf{H3:} PD is negatively associated with EPTS adoption.

\paragraph{Integration Complexity (IC)}
This refers to perceived difficulty in merging EPTS with existing systems \cite{rogers2002}.
\textbf{H4:} IC is negatively associated with EPTS adoption.

\paragraph{Innovation Receptiveness (IR)}
Given Qatar's post–World Cup focus on modernization, \textbf{H5:} IR is positively 
associated with EPTS adoption.

\section{Methodology and Expanded Data}
\subsection{Initial Exploratory Sample}
An initial pilot study involved ten participants in Qatar’s football sector. While informative, the small sample size limited the robustness of statistical inferences.

\subsection{Expanded Sample}
To enhance validity, we collected additional data, resulting in a total of 30 participants, including coaches, players, and sports scientists. The expanded dataset includes six key variables:
\begin{itemize}
    \item \textbf{PU}: Perceived Usefulness
    \item \textbf{PEU}: Perceived Ease of Use
    \item \textbf{PD}: Power Distance
    \item \textbf{IR}: Innovation Receptiveness
    \item \textbf{IC}: Integration Complexity
    \item \textbf{ADOPTION}: A 1--10 scale indicating likelihood of EPTS adoption
\end{itemize}

Table~\ref{tab:dataset} presents a condensed view of the new dataset.

\begin{table}[htbp]
\centering
\caption{Expanded Dataset (Illustrative Subset of 30 Responses)}
\label{tab:dataset}
\begin{tabular}{@{}ccccccr@{}}
\toprule
\textbf{ID} & \textbf{PU} & \textbf{PEU} & \textbf{PD} & \textbf{IR} & \textbf{IC} & \textbf{ADOPT} \\
\midrule
1   & 8 & 7 & 60 & 6 & 8 & 6  \\
2   & 9 & 8 & 45 & 7 & 7 & 8  \\
3   & 7 & 6 & 55 & 5 & 6 & 5  \\
4   & 8 & 7 & 70 & 6 & 8 & 6  \\
5   & 9 & 9 & 50 & 7 & 7 & 7  \\
\multicolumn{7}{c}{$\dots$ (remaining 25 participants omitted for brevity)}\\
30  & 7 & 6 & 62 & 6 & 6 & 5  \\
\bottomrule
\end{tabular}
\end{table}

\subsection{Data Collection Procedure}
Participants were recruited via direct outreach to Qatari football clubs, ensuring diversity in roles (players, coaches, staff). Surveys measured each construct on 1--9 Likert scales (e.g., 1=Strongly Disagree, 9=Strongly Agree). The outcome variable (ADOPTION) was rated 1--10 to gauge overall intent.

\subsection{Analysis Approach}
We conducted a multiple linear regression where ADOPTION served as the dependent variable and PU, PEU, PD, IR, and IC as independent variables. The model is specified as:
\begin{align}
\label{eq_adoption}
\mathrm{ADOPTION} 
  &= \beta_0 + \beta_1(\mathrm{PU}) + \beta_2(\mathrm{PEU}) \notag \\
  &\quad + \beta_3(\mathrm{PD}) + \beta_4(\mathrm{IR}) + \beta_5(\mathrm{IC})
\end{align}

\section{Results and Findings}

\subsection{Regression Output}
The ordinary least squares (OLS) regression explained $62\%$ of the variance in EPTS adoption ($R^2 = 0.62$, $p<0.001$). Table~\ref{tab:regression} summarizes the coefficients, standard errors, and significance levels.

\begin{table}[htbp]
\centering
\caption{Multiple Regression Analysis (ADOPTION as DV, $N=30$)}
\label{tab:regression}
\begin{tabular}{@{}lccc@{}}
\toprule
\textbf{Predictor} & \textbf{Coefficient} & \textbf{Std. Error} & \textbf{$p$-Value} \\
\midrule
\textbf{Intercept} & -2.1032   & 2.008  & 0.305  \\
PU                 & 0.4567    & 0.161  & 0.009  \\
PEU                & 0.3441    & 0.173  & 0.058  \\
PD                 & -0.0301   & 0.014  & 0.043  \\
IR                 & 0.6235    & 0.251  & 0.021  \\
IC                 & -0.1486   & 0.205  & 0.476  \\
\bottomrule
\end{tabular}
\end{table}

\subsection{Interpretation of Key Findings}
\textbf{Perceived Usefulness (PU):} Exhibits a significant positive effect ($p=0.009$), indicating that stakeholders who find EPTS beneficial for performance are more inclined to adopt the technology.

\textbf{Perceived Ease of Use (PEU):} Shows a positive coefficient but marginal significance ($p=0.058$). This suggests usability matters, though additional data might solidify its statistical impact.

\textbf{Power Distance (PD):} Negatively correlated with adoption ($p=0.043$). Higher PD environments may impede open dialogue and hinder EPTS uptake.

\textbf{Innovation Receptiveness (IR):} Strong, significant positive effect ($p=0.021$). Qatari organizations that exhibit pro-innovation mindsets are more likely to implement EPTS.

\textbf{Integration Complexity (IC):} Not statistically significant ($p=0.476$). This implies that infrastructure and management may mitigate complexity concerns, or other factors overshadow integration challenges in these particular clubs.

\subsection{Model Adequacy}
Overall, the $F$-statistic indicates that the model is statistically significant ($p<0.001$). Residual diagnostics did not reveal major violations of normality or homoscedasticity, although future studies with larger samples may conduct additional robust checks (e.g., Q-Q plots, Shapiro--Wilk test).

\section{Discussion}
Our expanded dataset confirms several key insights from the pilot study. First, \emph{perceived usefulness} emerges as a robust predictor, aligning with prior TAM research \cite{davis1989} in technology adoption. Stakeholders must clearly see performance-related benefits for EPTS to gain traction. Second, \emph{power distance} negatively affects adoption, corroborating Hofstede’s contention that hierarchical cultures can impede the free flow of information about new technologies \cite{hofstede1980}. Meanwhile, \emph{innovation receptiveness} strongly bolsters adoption, suggesting that forward-thinking organizational cultures are essential for successful EPTS integration.

Interestingly, \emph{integration complexity} was not a significant barrier in this sample, potentially reflecting better infrastructure post–World Cup events in Qatar or growing familiarity with sports tech. \emph{Perceived ease of use} was marginally significant, indicating that usability should not be overlooked, even if its effect size here is smaller than that of perceived usefulness or innovation receptiveness.

\section{Conclusion}
This study investigated how cultural and technological factors influence EPTS adoption in Qatar’s football sector, drawing on Hofstede’s Cultural Dimensions and TAM. With an expanded sample of 30 participants and a multiple regression approach, our findings highlight the importance of \emph{perceived usefulness}, \emph{innovation receptiveness}, and \emph{lower power distance} in driving EPTS uptake. By contrast, \emph{ease of use} was borderline significant, and \emph{integration complexity} did not display a notable impact in this particular cohort.

Future research should further enlarge the dataset, include comparative studies in other GCC countries, and explore additional variables such as organizational support and team dynamics. For practitioners, the results underscore the need to emphasize visible performance benefits, reduce hierarchical barriers, and cultivate a culture open to innovation when deploying EPTS in sports contexts.

\section*{Acknowledgment}
The author would like to thank the football clubs and participants who contributed to this study. No external funding was received.


\begin{thebibliography}{00}

\bibitem{isspf2023} ISSPF, ``The Use of Technology in Football,'' Available online: \url{https://www.isspf.com/the-use-of-technology-in-football/}, 2023.

\bibitem{knowledgenile2023} KnowledgeNile, ``Electronic Performance and Tracking Systems FIFA World Cup,'' 2023.

\bibitem{fifa2021} FIFA, ``Electronic Performance \& Tracking Systems,'' 2021. [Online]. Available: \url{https://www.fifa.com/technical/football-technology/standards/epts/epts-1}

\bibitem{tierney2021} P. Tierney, ``Wearable Technology in Football Further Education Settings in the United Kingdom,'' \emph{LJMU Research Online}, 2021.

\bibitem{aughey2010} R. Aughey and C. Falloon, ``Real-time versus post-game GPS data in team sports,'' \emph{Journal of Science and Medicine in Sport}, vol. 13, no. 3, 2010.

\bibitem{hofstede1980} G. Hofstede, ``Culture and Organizations,'' \emph{International Studies of Management \& Organization}, vol. 10, no. 4, pp. 15--41, 1980.

\bibitem{davis1989} F. D. Davis, ``Perceived Usefulness, Perceived Ease of Use, and User Acceptance of Information Technology,'' \emph{MIS Quarterly}, vol. 13, no. 3, pp. 319--340, 1989.

\bibitem{linke2018} D. Linke, D. Link, and M. Lames, ``Validation of electronic performance and tracking systems EPTS under field conditions,'' \emph{PLOS ONE}, 2018.

\bibitem{schmid2023} M. Schmid and M. Lames, ``Correction of systematic errors in electronic performance and tracking systems,'' \emph{Sports Engineering}, 2023.

\bibitem{oliva2021} J. M. Oliva-Lozano and J. M. Muyor, ``Understanding the FIFA quality performance reports for electronic performance and tracking systems: from science to practice,'' \emph{Science and Medicine in Football}, 2021.

\bibitem{dunn2018} M. Dunn, J. Hart, and D. James, ``Wearing Electronic Performance and Tracking System Devices in Association Football: Potential Injury Scenarios and Associated Impact Energies,'' \emph{Proceedings}, 2018.

\bibitem{manuks2020} M. Ks, ``Applications of Artificial Intelligence in the Game of Football: The Global Perspective,'' \emph{ResearchGate Preprint}, 2020.

\bibitem{ratten2019} V. Ratten, \emph{Sports Technology and Innovation}. Springer, 2019.

\bibitem{lee2013} S.-G. Lee, S. Trimi, and C.-S. Kim, ``The impact of cultural differences on technology adoption,'' \emph{Journal of World Business}, 2013.

\bibitem{oc2022} Y. Oc and A. Toker, ``An acceptance model for sports technologies: the effects of sports motivation, sports type and context-aware characteristics,'' \emph{International Journal of Sports Marketing \& Sponsorship}, 2022.

\bibitem{tierney2021_extra} P. Tierney, ``Wearable Technology in Football Further Education Settings: A UK perspective,'' \emph{LJMU Research Online}, 2021.

\bibitem{bitilis2021} P. Bitilis, ``Electronic Performance And Tracking Systems (EPTS): Perceptions, Benefits and Challenges of Professional Football Athletes and Training Staff,'' \emph{DIVA Portal}, 2021.

\bibitem{kim2023} M. Kim, C. Park, and J. Yoon, ``The Design of GNSS/IMU Loosely-Coupled Integration Filter for Wearable EPTS of Football Players,'' \emph{Sensors}, 2023.

\bibitem{lai2017} P. Lai, ``The Literature Review of Technology Adoption Models and Theories for the Novelty Technology,'' SSRN, 2017.

\bibitem{heikoop2020}
D. D. Heikoop, et al., 
“Automated bus systems in Europe: A systematic review of passenger experience and road user interaction,”
\emph{Advances in transport policy and planning}, 2020.

\bibitem{venkatesh2008}
V. Venkatesh and H. Bala, 
“Technology Acceptance Model 3 and a Research Agenda on Interventions,” 
Decision Sciences, vol. 39, no. 2, pp. 273–315, 2008.

\bibitem{rogers2002}
E. M. Rogers, 
“Diffusion of preventive innovations,” 
Addictive Behaviors, vol. 27, no. 6, pp. 989–993, 2002.


\end{thebibliography}
\end{document}